\documentclass[12pt]{article}

\begin{document}

\thispagestyle{empty}
\renewcommand{\thefootnote}{\fnsymbol{footnote}}

\begin{flushright}
{\small
SLAC--PUB--8589\\
August 2000\\}
\end{flushright}

\vspace{.8cm}

\begin{center}
{\bf\large   
Simulation of the Beam-Beam Effects in $e^+e^-$ Storage Rings with a 
Method of Reducing the Region of Mesh\footnote{Work supported by 
Department of Energy contract DE--AC03--76SF00515.}}

\vspace{1cm}

Yunhai Cai, Alex W. Chao and Stephan I. Tzenov\\
Stanford Linear Accelerator Center, Stanford University,
Stanford, CA 94309\\

\medskip

Toshi Tajima\\
University of Texas at Austin, Austin, TX 78712 \\
and Lawrence Livermore National Laboratory, Livermore, CA 94551\\
\end{center}

\vfill

\begin{center}
{\bf\large   
Abstract }
\end{center}

\begin{quote}
A highly accurate self-consistent particle code to simulate the beam-beam
collision in $e^+e^-$ storage rings has been developed. It adopts a method 
of solving the Poisson equation with an open boundary. The method consists 
of two steps: assigning the potential on a finite boundary using the 
Green's function, and then solving the potential inside the boundary with 
a fast Poisson solver. Since the solution of the Poisson's equation is 
unique, our solution is exactly the same as the one obtained by simply 
using the Green's function. The method allows us to select much smaller 
region of mesh and therefore increase the resolution of the solver. The 
better resolution makes more accurate the calculation of the dynamics in 
the core of the beams. The luminosity simulated with this method agrees 
quantitatively with the measurement for the PEP-II B-factory ring in the 
linear and nonlinear beam current regimes, demonstrating its 
predictive capability in detail. 

\end{quote}

\vfill

\begin{center} 


{\it Submitted to Physical Review Special Topics: Accelerators and Beams}
\end{center}

\newpage



%
\pagestyle{plain}

\renewcommand{\theequation}{\thesection.\arabic{equation}}

\setcounter{equation}{0}

\section{Introduction}

The beam-beam interaction is one of the most important limiting factors 
determining the luminosity of storage colliders. It has been studied 
extensively by theoretical analysis \cite{chao}, experimental measurements 
\cite{seeman}, and computer simulations \cite{myers}. Historically, due 
to the complexity of the interaction, many approximations, such as 
strong-weak \cite{hirata} or soft-Gaussian \cite{furman}, have been 
introduced in order to simulate the interaction in a reasonable computing 
time. The self-consistent simulation of the beam-beam interaction by solving 
the Poisson equation with a boundary condition has been proposed first to 
investigate the round beams \cite{siemann} and then the flat beams 
\cite{krishnagopal}. To enhance the accuracy and to reduce the 
computational overhead, an algorithm (and a code) of the so-called 
$\delta f$ method that can handle strong-strong interactions has been 
introduced \cite{koga}. Another self-consistent approach to the beam-beam 
interaction is to use the Green's function directly \cite{anderson} or 
indirectly \cite{ohmi}. 

In the present paper we will develop a method that takes advantage 
from both self-consistent approaches: a smaller region of mesh from the 
method of using the Green's function and a faster solver for the interior. 
In order to develop a highly accurate predictive code at the luminosity 
saturation region, it is necessary to have a fully self-consistent treatment 
of field-particle interaction at collision. Since we are interested in 
simulating the Asymmetric $e^+ e^-$ Storage Collider PEP-II \cite{book}, 
which needs to maximize the luminosity and thus the beam current, it is 
even more crucial that the beam-beam interaction in the large current 
regime be treated accurately.

In a self-consistent simulation of the beam-beam interaction in storage 
rings, the beam distributions have to be evolved dynamically during 
collision with the opposing beam together with the propagation in the 
rings. During collision, the beam distributions are used at each time 
sequence to compute the force that acts on the opposing beam.

Since positrons and electrons are ultra-relativistic particles in high 
energy storage rings, the beam-beam force is transverse and acts only 
on the opposing beam. Hence, given a beam distribution, we can divide the 
distribution longitudinally into several slices and then solve for 
the two-dimensional force for each slice. Self-consistency is achieved
by introducing many-body particles in the field that in turn constitute 
charge-current, the strategy of the particle-in-cell (PIC) procedure 
(for example, Ref. \cite{tajima}). In this paper, for simplicity, we use 
only a single longitudinal slice for a bunch, ignoring any beam-beam effects 
encompassing over the length of the bunch.

\renewcommand{\theequation}{\thesection.\arabic{equation}}

\setcounter{equation}{0}

\section{Method}

In modern colliders, beams are focused strongly at the interaction point 
to achieve high luminosity. As a result the transverse dimension of the
beam is much smaller than the dimension of the beam pipe at the collision 
point. Therefore, the open boundary condition is a good approximation 
for calculating the transverse beam-beam force.

\subsection{Green's Function}

Given a charge density $\rho_c(x,y)$, which is normalized to the total charge

\begin{equation}
\int dx dy \rho_c {\left( x, y \right)} = Ne, 
\end{equation}

\noindent
where $N$ is the total number of particles, the electric potential $\phi(x,y)$
satisfies the Poisson equation

\begin{equation}
{\left( {\partial^2\over\partial x^2} + 
{\partial^2\over\partial y^2} \right)} \phi 
{\left( x, y \right)} = - 2 \pi \rho_c 
{\left( x, y \right)} 
\end{equation} 

\noindent
with $x$ and $y$ being the transverse coordinates. The solution of the 
Poisson equation can be expressed as 

\begin{equation}
\phi {\left( x, y \right)} = 
\int dx^\prime dy^\prime G {\left( x-x^\prime, \; 
y-y^\prime \right)} \rho_c 
{\left( x^\prime, y^\prime \right)}, 
\label{eqn:green}
\end{equation}

\noindent
where $G$ is the Green's function which satisfies the equation 

\begin{equation}
{\left( {\partial^2\over\partial x^2} + 
{\partial^2\over\partial y^2} \right)} 
G {\left( x-x^\prime, \; y-y^\prime \right)} = 
- 2 \pi \delta {\left( x-x^\prime \right)} 
\delta {\left( y-y^\prime \right)}. 
\end{equation}

\noindent
In the case of open boundary condition, namely the boundary is far away 
so that its contribution to the potential can be ignored, one has the 
well-known explicit solution for the Green's function: 

\begin{equation}
G {\left( x-x^\prime, y-y\prime \right)} = - 
{1\over2} \ln {\left[ {\left( x-x^\prime \right)}^2 + 
{\left( y-y^\prime \right)}^2 \right]}. 
\end{equation}

\noindent
This explicit solution can be used directly to compute the potential. The
main problem of this approach is that it is slow to calculate the logarithm 
and the number of computations is proportional to the square of the number 
of macro particles $N_p^2$. One can reduce $N_p$ by introducing a 
two-dimensional mesh to smooth out the charge distribution \cite{anderson}. 
Or to further improve the computing speed, one can map the solution onto 
the space of spectrum by the Fast Fourier Transformation (FFT) and then 
calculate the potential \cite{ohmi}.  

\subsection{Reduce the Region of Mesh}

Another alternative approach is to solve the Poisson equation with a 
boundary condition \cite{krishnagopal}, because the region 
(20 $\mu$m $\times$ 450 $\mu$m for PEP-II) occupied by the beam is much 
smaller than the boundary defined by the beam pipe (2 cm radius) at the 
collision point. In order to achieve required resolution, a few mesh 
points per $\sigma$ of the beam are needed, otherwise the size of mesh 
is too large for numerical computation.

However, it is unnecessary to cover the entire area with mesh inside the 
beam pipe since the area is mostly empty. We choose a smaller and finite
area of the mesh, which is large enough to cover the whole beam, and by 
carefully selecting the potential on the boundary, we can obtain the 
accurate solution inside the boundary. 

We denote by $\phi_1$ the solution (\ref{eqn:green}) of the Poisson equation. 
Let $\phi_2$ be the solution obtained by solving the Poisson equation in a 
two-dimensional area $S$ with the potential prescribed on a closed 
one-dimensional $L$ bounding the area $S$ 

\begin{equation}
\phi_2 {\left(x, y \right)} = \int \limits_S dx^\prime 
dy^\prime G {\left( x-x^\prime, \; 
y-y^\prime \right)} \rho_c 
{\left( x^\prime, y^\prime \right)}, 
\end{equation}

\noindent
where $(x, y) \in L$. By definition, we have $\phi_1 = \phi_2$ on the 
boundary $L$. Let $U = \phi_1 - \phi_2$ and use the first identity of 
Green's theorem \cite{jackson} in two dimensions 

\begin{equation}
\int \limits_S {\left[ U \nabla^2 U + 
{\left( \nabla U \right)}^2 \right]} dx dy = 
\oint \limits_L U {\partial U\over\partial n} dl, 
\end{equation} 

\noindent
where $dl$ is a line element of $L$ with a unit outward normal $n$. Since
$U=0$ on $L$ and $\nabla^2U = 0$ inside $L$, we have

\begin{equation}
\int \limits_S {\left( \nabla U \right)}^2 dxdy = 0, 
\end{equation}

\noindent
implying that $U$ is a constant inside $L$. We can set $U = 0$, which is
consistent with the value on the boundary. Hence $\phi_1 = \phi_2$. The
two solutions are identical.

\renewcommand{\theequation}{\thesection.\arabic{equation}}

\setcounter{equation}{0}

\section{Field Solver}

We adopt the PIC technique to calculate the fields induced by 
the charge (and current) of the beams self-consistently. 
The charge distribution of a beam is represented by macro particles. These
macro particles are treated as single electron or positron dynamically. In
order to compute the field acting on the particles of the opposing beam,
we first deposit their charges onto the gird points of a two-dimensional 
rectangular mesh. We denote by $H_x$ the horizontal distance between two
adjacent grid points and by $H_y$ the distance in vertical direction.  

\subsection{Charge Assignment}

\begin{figure}[ht]
 \vspace{9.0cm}
\includegraphics{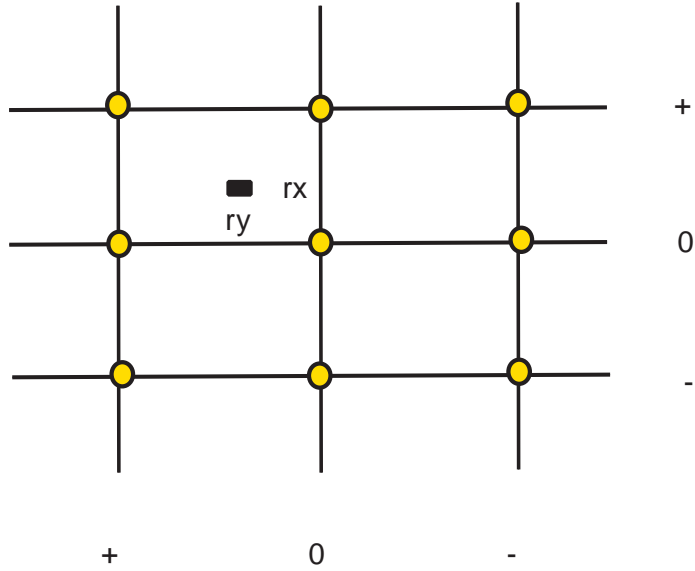}
 \caption{\it Scheme of charge assignment.
    \label{fig:weight} }
\end{figure}

We choose the method of the triangular-shaped cloud \cite{eastwood}
as our scheme for the charge assignment onto the grid. On a two-dimensional 
grid, associated with each macro particle, nine nearest points are 
assigned with non-vanishing weights as illustrated in Fig. ~\ref{fig:weight}. 
We use ``0'' to denote the first, ``+'' as the second, and ``-'' as the 
third nearest lines.

The weights are quadratic polynomials of the fractional distance,
$r_x = {\delta x / H_x}$, to the nearest line

\begin{eqnarray}
w_x^0 &=& {3\over4} - r_x^2, \nonumber \\ 
w_x^+ &=& {1\over2} {\left( {1\over4} + r_x + 
r_x^2 \right)}, \\ 
w_x^- &=& {1\over2} {\left( {1\over4} - r_x + 
r_x^2 \right)}. \nonumber 
\end{eqnarray}

\noindent
The coefficients are chosen such that the transition at the middle
of the grid is continuous and smooth, and $w_x^0 + w_x^+ + w_x^- = 1$ 
which is required by the conservation of charge. In order to retain 
these properties, the weights of the two-dimensional grid are simply 
a product of two one-dimensional weights. For example, 
$w^{00} = w_x^0 w_y^0$ or $w^{+-} = w_x^+ w_y^-$.

\subsection{Poisson Solver}

It is crucial to solve the Poisson equation fast enough (within a second on
a computer workstation) for the beam-beam simulation, because the radiation 
damping time is about 5000 turns and several damping times are needed to 
reach an equilibrium distribution. For the reason of the computing speed, we
follow Krishnagopal \cite{krishnagopal} and choose the method of cyclic 
reduction and FFT \cite{hockney}. A five-point difference scheme is used to 
approximate the two-dimensional Laplacian operator

\begin{equation}
{\phi_{i-1,j} + \phi_{i+1,j} -2\phi_{i,j} \over H_x^2}
+ {\phi_{i,j-1} + \phi_{i,j+1} -2\phi_{i,j} \over H_y^2}
= -2 \pi {\rho_c}_{i,j}, 
\end{equation}

\noindent
where $i$ and $j$ are the horizontal and vertical indices that label the 
grid points on the mesh.

Truncation errors are of the order of $H_x^2$ and $H_y^2$. It is worthwhile 
to mention that, if we use the same number of mesh points per $\sigma$ in 
both transverse directions in the case of beam aspect ratio 30:1,  the 
truncation errors in the horizontal plane are dominant. To minimize the 
errors in our simulation, we select three times more mesh points per 
$\sigma$ in horizontal direction compared to the vertical one.

\subsection{Field}

The field ${\vec E} = -\nabla\phi$ is computed on the two dimensional
grid, using a six-point difference scheme

\begin{eqnarray}
{E_x}_{i,j} & = & - {1\over 12 H_x} 
{\left[ {\left( \phi_{i+1,j+1} - \phi_{i-1,j+1} \right)} 
+ 4 {\left( \phi_{i+1,j} - \phi_{i-1,j} \right)} 
+ {\left( \phi_{i+1,j-1} - \phi_{i-1,j-1} 
\right)} \right]}, \\ 
{E_y}_{i,j} & = & - {1\over 12 H_y}
{\left[ {\left( \phi_{i+1,j+1} - \phi_{i+1,j-1} \right)} 
+ 4 {\left( \phi_{i,j+1} - \phi_{i,j-1} \right)} 
+ {\left( \phi_{i-1,j+1} - \phi_{i-1,j-1} 
\right)} \right]}.
\end{eqnarray}

\noindent
The field off the grid is computed with the same smoothing scheme used
in the charge assignment to ensure the conservation of the momentum. The 
fields $E_x$ and $E_y$ are interpolated between the grid points. They 
are calculated by using the weighted summation of the fields at the nine 
nearest points with exactly the same weights used as the charge is 
assigned. 

\renewcommand{\theequation}{\thesection.\arabic{equation}}

\setcounter{equation}{0}

\section{Track Particles}

The motion of a particle is described by its canonical coordinates

\begin{equation}
z^T = {\left( x, P_x, y, P_y \right)}, 
\end{equation}

\noindent
where $P_x$ and $P_y$ are particle momenta normalized by the design 
momentum $p_0$. 

\subsection{One-Turn Map}

When synchrotron radiation is turned off, a matrix is used to describe 
the linear motion in the lattice 

\begin{equation}
z_{n+1} = M \cdot z_n, 
\end{equation}

\noindent
where $M$ is a $4\times4$ symplectic matrix which can be partitioned into 
blocks of $2\times2$ matrices when the linear coupling is ignored 

\begin{equation}
  M = \left( \begin{array}{ll}
             M_x & 0 \\
             0 & M_y  
             \end{array}
      \right). 
\end{equation}

\noindent
Here $M_x$, and $M_y$ are $2\times2$ symplectic matrices. The 
matrix $M_x$ is expressed with the Courant-Snyder parameters $\beta_x$, 
$\alpha_x$, and $\gamma_x$ at the collision point  

\begin{equation}
  M_x = \left(  \begin{array}{ll} 
   \cos(2\pi\nu_x)+\alpha_x \sin(2\pi\nu_x) & \beta_x \sin(2\pi\nu_x) \\
   -\gamma_x \sin(2\pi\nu_x) & \cos(2\pi\nu_x)-\alpha_x \sin(2\pi\nu_x) 
                 \end{array} 
        \right), 
\end{equation}

\noindent
where $\nu_x$ is the horizontal tune. A similar expression is applied in 
the vertical plane. 

\subsection{Damping and Synchrotron Radiation}

Following Hirata \cite{ruggiero}, we apply the radiation damping and 
quantum excitation in the normalized coordinates, since it is easily 
generalized to include the linear coupling. The motion of a particle in 
the normalized coordinate is described by a rotation matrix

\begin{equation} 
   R_x = \left(  \begin{array}{ll}
                         \cos(2\pi\nu_x)  & \sin(2\pi\nu_x)  \\
                         -\sin(2\pi\nu_x) & \cos(2\pi\nu_x) 
                            \end{array} 
                    \right), 
\end{equation}

\noindent
which is obtained by performing the similarity transformation

\begin{equation}
   R_x = A_x^{-1} \cdot M_x \cdot A_x, 
\end{equation}

\noindent
where 

\begin{equation}
  A_x =  \left(  \begin{array}{ll}
                   {\sqrt\beta_x}            & 0 \\
                  -{\alpha_x\over\sqrt\beta_x} & {1\over\sqrt\beta_x}
              \end{array} \right),
  A_x^{-1} = \left( \begin{array}{ll}
                   {1\over\sqrt\beta_x}      & 0 \\
                   {\alpha_x\over\sqrt\beta_x} & {\sqrt\beta_x}
                    \end{array} 
             \right).
\end{equation}

When synchrotron radiation is switched on, we simply replace the rotation 
matrix $R_x$ with following map in the normalized coordinates ${\bar x}$ and 
${\bar P_x}$

\begin{equation}
   \left( \begin{array}{l} 
            {\bar x} \\ 
            {\bar P_x} 
          \end{array} 
   \right)
           = e^{-{1\over\tau_x}} R_x
   \left( \begin{array}{l} 
            {\bar x} \\ 
            {\bar P_x} 
          \end{array} 
   \right)
           + \sqrt{\epsilon_x(1-e^{-{2\over\tau_x}})}
   \left( \begin{array}{l} 
            {\eta_{\bar x}} \\ 
            {\eta_{\bar P_x}} 
          \end{array} 
   \right),
\end{equation}

\noindent
where $\eta_{\bar x}$ and $\eta_{\bar p_x}$ are Gaussian random variables 
normalized to unity, $\tau_x$ is the damping time in unit of number 
of turns and $\epsilon_x$ is the equilibrium emittance. In the vertical 
plane, a similar map is applied.

\subsection{Beam-Beam Kick}

Assuming particles are ultra-relativistic and the collision is head-on, 
the kick on a particle by the opposing beam is given by the Lorenz force

\begin{eqnarray}
  \delta P_x &=& - {2e\over E_0} E_x, \\
  \delta P_y &=& - {2e\over E_0} E_y, 
\end{eqnarray}

\noindent
where $E_x$ and $E_y$ are the horizontal and vertical components of the
electric field evaluated at the position of the particle. They are computed 
with the Poisson solver as outlined in the previous section each time 
two slices of the beam pass each other. And the half of the transverse
force is the magnetic force due the beam moving at the speed of light.
The energy of the particle, $E_0 = cp_0$, appearing in the denominator of 
the above expressions comes from the normalization of the canonical 
momenta $P_x$ and $P_y$ and the use of the s-coordinate, $s=ct$, as the 
``time'' variable.   

\begin{figure}[ht]
 \vspace{9.0cm}
\includegraphics{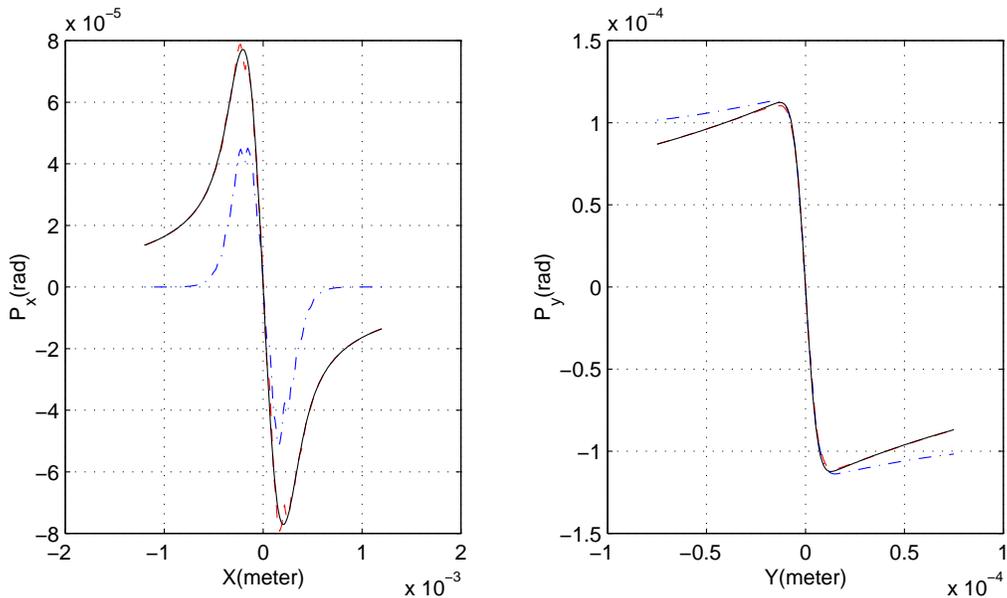}
 \caption{\it The beam-beam kick by a flat Gaussian beam with aspect ratio 
 30:1 near X axis and Y axis. The dash-dotted curve is the case when 
$\phi = 0$ is assigned as the boundary condition. The long-dashed curve is 
the kick when inhomogeneous boundary condition is used. The short-dashed 
curve is the kick produced by the Erskine-Bassetti formula \cite{bassetti}. 
    \label{fig:kick} }
\end{figure}

A typical beam-beam kick experienced by a particle near the axis is shown
in Fig.~\ref{fig:kick} with the PEP-II parameters, which are tabulated
in the next section.  As expected based on the derivation in section 2.2, 
the kick resulted from solving the Poisson equation with the inhomogeneous 
boundary condition agrees well with the analytic solution. In addition,
the agreement demonstrates that the scheme of the charge deposition works
well, the mesh is dense enough and the number of macro particles is large
enough. 

The number of macro particles used to represent the distribution of the 
beam is 10240. The area of the mesh is $8\sigma_x$$\times$$24\sigma_y$ and 
there are 15 grid points per $\sigma_x$ and 5 per $\sigma_y$. 
There are about 15 macro particles per cell within 3$\sigma$ of the beam. 
These parameters are chosen to minimize truncation errors and maximize 
resolution. The 256$\times$256 mesh is also the maximum allowed by a 
computer workstation to complete a typical job within a reasonable time.

The discrepancy between the solution with the homogeneous boundary condition, 
$\phi = 0$, and the analytic one worsen as the beam aspect ratio becomes 
larger because the actual change of the potential on the horizontal 
boundaries becomes larger.

\renewcommand{\theequation}{\thesection.\arabic{equation}}

\setcounter{equation}{0}

\section{Simulation of PEP-II: Validation}

An object-oriented C++ class library has been written to simulate the 
beam-beam interaction using the method outlined in the previous sections. 
In the library, the beam and the Poisson solver are all independent objects 
that can be constructed by the user. For example, there is no limitation 
on how many beam objects are allowed in the simulation and the beams can 
have different parameters as an instance of the beam class. These 
features provide us with great flexibility to study various phenomena of 
the beam-beam interaction.

We will carry out the simulation of beam-beam interaction with the current 
operating parameters of the PEP-II so that the results of the simulation 
can be compared with the known experimental observations. As a goal of this 
study, after a proper benchmarking of the code against the experiment, we 
shall be able to make predictions on parameter dependence and show how to 
improve the luminosity performance of the collider.
  
\subsection{PEP-II Operating Parameters}

\begin{table}[h]
\begin{center}
\begin{tabular}{llll}
\hline
\hline
Parameter               & Description              & LER(e+)            
& HER(e-)            \\ \hline
$E$ (Gev)               & Beam energy              & 3.1                
& 9.0                \\
$\beta_x^*$ (cm)        & Beta X at the IP         & 50.0               
& 50.0               \\
$\beta_y^*$ (cm)        & Beta Y at the IP         & 1.25               
& 1.25               \\
$\tau_t$ (turn)         & Transverse damping time  & 9740               
& 5014               \\
$\epsilon_x$ (nm-rad)   & Emittance X              & 24.0               
& 48.0               \\
$\epsilon_y$ (nm-rad)   & Emittance Y              & 1.50               
& 1.50               \\
$\nu_x$                 & X tune                   & 0.649              
& 0.569              \\
$\nu_y$                 & Y tune                   & 0.564              
& 0.639              \\
  
\hline
\hline
\end{tabular}
\end{center}
\label{tab:pepii}
\vskip 10pt
\hskip 106pt Table 5.1: {\it Parameters for the beam-beam simulation}  
\end{table}

The parameters used in the simulation are tabulated in Tab.~\ref{tab:pepii}.
The vertical $\beta_y^*$ is lowered to 1.25cm \cite{yuri} from the design 
value 1.5cm \cite{book}. The horizontal emittance 24nm-rad in the Low Energy 
Ring (LER) is half of the design value 48nm-rad because the wiggler is turned 
off to increase the luminosity. The damping time, 9740 turns, in the LER is a 
factor of two larger than the one in the High Energy Ring (HER) because of 
the change of the wigglers made during the construction of the machine. The 
degradation of luminosity from the increase of the damping time was found 
then to be about 10\% based on the beam-beam simulation. The tunes are split 
and are determined experimentally to optimize the peak luminosity.

\subsection{Procedure of simulation}

The distribution of the beam is represented as a collection of macro 
particles that are dynamically tracked. The procedure to obtain equilibrium 
distributions of the two colliding beams is as follows 

\vspace{0.2cm}

\indent $\bullet$ initialize the four-dimensional Gaussian distribution
                  according to the parameters of the lattice at the collision
                  point and the emittance of the beam.  
                  Distributions of two beams are independent and different.

\indent $\bullet$ iterate a loop with three damping times 

\indent $\bullet$ propagate each beam through corresponding lattice
                  using one-turn map with synchrotron radiation.

\indent $\bullet$ cast the particle distributions onto the grid
                  as the charge distribution with weighting and smoothing.

\indent $\bullet$ solve for the potential on the grid with the Poisson 
                  solver.

\indent $\bullet$ compute the field on the grid.

\indent $\bullet$ calculate the beam-beam kick to the particles of the
                  other beam with the field at the position of the particles.
                  The field off the grid is interpolated with the same
                  weighting and smoothing used in the charge deposition.

\indent $\bullet$ save data such as beam size, beam centroid and luminosity.

\indent $\bullet$ end of the loop.

\indent $\bullet$ save the final distributions.

\vspace{0.2cm} 

We vary the beam intensity with a fixed beam current ratio: $I_+$:$I_- =
$ 2:1, which is close to the ratio for the PEP-II operation. At each beam 
current, we compute the equilibrium distributions. 

\subsection{Beam-Beam Limit}

Given equilibrium distributions that are close enough to the Gaussian, we can
introduce the beam-beam parameters

\begin{eqnarray}
 \xi_x^\pm = {r_e N^\mp\beta_x^\pm \over 
              2\pi\gamma^\pm\sigma_x^\mp(\sigma_x^\mp + \sigma_y^\mp) }, 
             \nonumber \\
 \xi_y^\pm = {r_e N^\mp\beta_y^\pm \over 
              2\pi\gamma^\pm\sigma_y^\mp(\sigma_x^\mp + \sigma_y^\mp) },
\end{eqnarray}

\noindent
where $r_e$ is the classical electron radius, $\gamma$ is the energy of 
the beam in unit of the rest energy, and $N$ is total number
of the charge in the bunch. Here the superscript ``+'' denotes quantities 
corresponding to the positron and ``-'' quantities corresponding to the 
electron. 

\begin{figure}[ht]
 \vspace{9.0cm}
\includegraphics{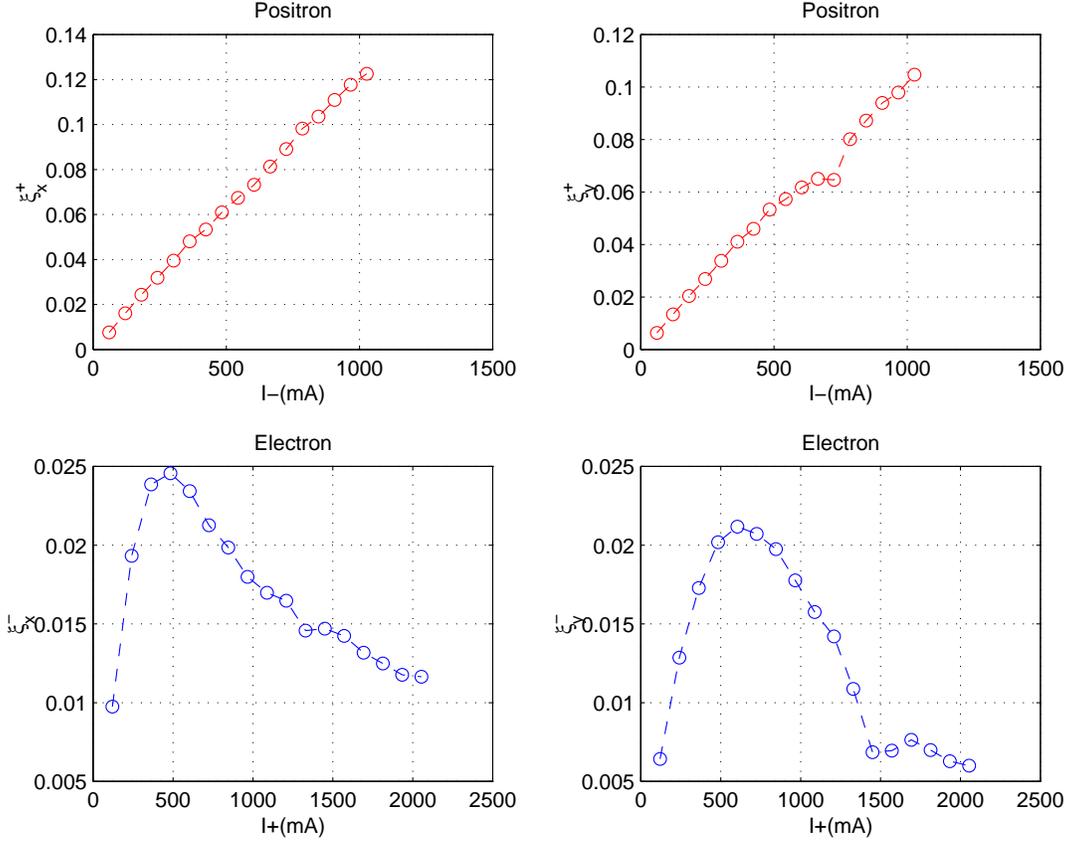}
 \vspace{3.5cm}
 \caption{\it The beam-beam tune shifts as a function of beam currents. 
    Number of bunches, $n_b = 554$, is used for the total beam currents. 
    The revolution frequency $f_0 = 136.312$ k$H_z$}. 
    \label{fig:shift} 
\end{figure}

The results of the simulation are shown in Fig.~\ref{fig:shift}.  
The beam-beam tune shifts for the electron beam are low because of the 
large beam-beam blowup of the positron beam. At this operating point, the 
positron is the weaker beam. When $I_+ = 1200$mA and $I_- = 600$mA, which
is the near the maximum allowed currents when the beams are in collision, 
the positron beam sizes are $\sigma_x^+ = 260\mu$m and 
$\sigma_y^+ = 7\mu$m. 
  
\subsection{Luminosity}

Given the two beam distributions, $\rho^+$ and $\rho^-$, the luminosity can 
be written as

\begin{equation}
L = n_b f_0 N^+ N^-\int \limits_{-\infty}^\infty 
\int \limits_{-\infty}^\infty 
\rho^+(x,y)\rho^-(x,y)dxdy, 
\end{equation}

\noindent
where $n_b$ is the number of the colliding bunches, $f_0$ is the revolution
frequency, and $N^+, N^-$ are the number of charges in each position
and electron bunch, respectively. Since the distribution $\rho$
is normalized to unity

\begin{equation}
\int dx dy \rho(x,y) = 1
\end{equation}

\noindent
and proportional to the charge density $\rho_c$, we evaluate the overlapping 
integral by a summation over $\rho_c^+\rho_c^-$ on the mesh. If we assume
the distributions are Gaussian, the overlapping integral can be carried out

\begin{equation}
L = {n_b f_0 N^+ N^- \over 2\pi\Sigma_x\Sigma_y}, 
\end{equation}

\noindent
where $\Sigma_x = \sqrt{{\sigma_x^+}^2 + {\sigma_x^-}^2}$ and  
$\Sigma_y = \sqrt{{\sigma_y^+}^2 + {\sigma_y^-}^2}$. Two methods agree within
a few percents. The mesh method gives a higher luminosity than the 
Gaussian one. We always use the mesh method, since it can be applied to broad 
classes of distribution.

\begin{figure}[ht]
 \vspace{9.0cm}
\includegraphics{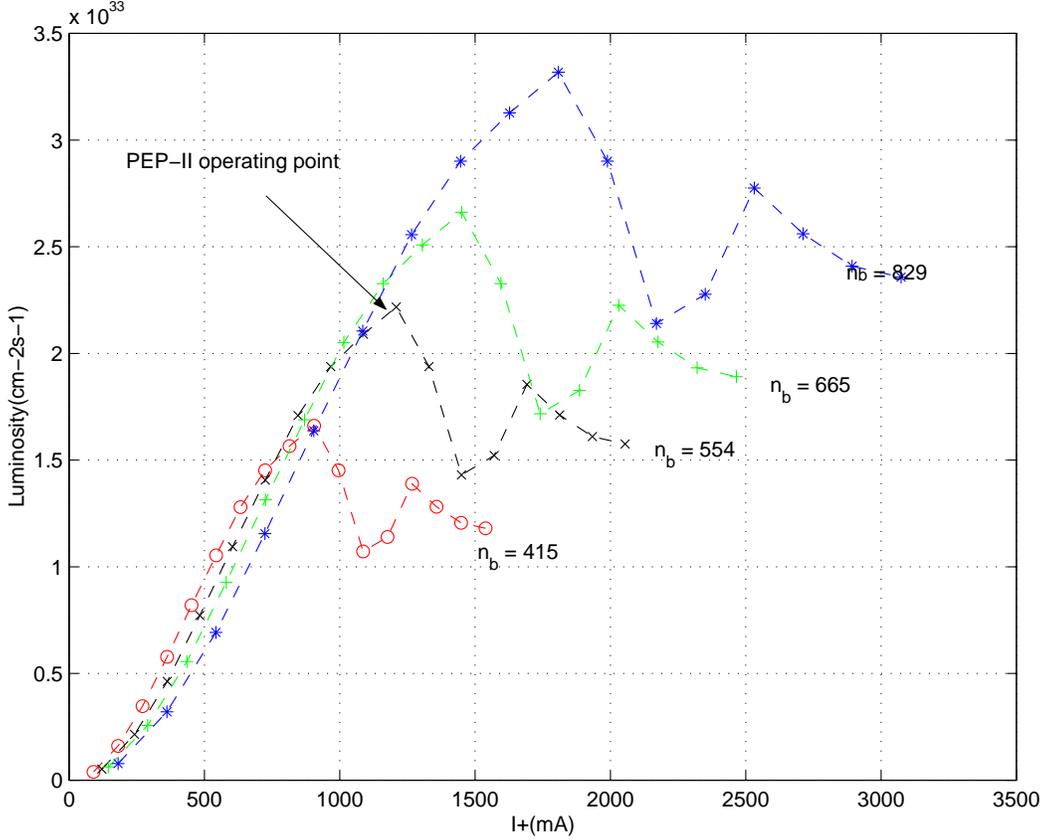}
 \vspace{3.5cm}
 \caption{\it Luminosity as a function of the beam current. The labels are
    the number of the colliding bunches.} 
    \label{fig:luminosity} 
\end{figure}

Figure~\ref{fig:luminosity} shows the luminosity of the beams 
with 415 colliding bunches, which are spaced with every 8 RF buckets and 
10\% of the gap. The luminosity is beam-beam limited. It also shows that 
the optimum number of bunches is between 544 and 665 and the luminosity is 
about 2.3$\times10^{33}{\rm cm}^{-2}{\rm s}^{-1}$ given $I_+ = 1200$mA. 
These results quantitatively agree with the experimental observations in the 
routine operation of the PEP-II. For example, the peak luminosity of the 
PEP-II is 1.95$\times10^{33}{\rm cm}^{-2}{\rm s}^{-1}$ with $I_+ = 1170$mA, 
$I_- = 700$mA, and $n_b = 665$ during the period of June, 2000. The fact 
that the luminosity value in the simulation is higher than the observation 
could be explained by the hour-glass effect which is ignored in the 
simulation.

For a fixed number of bunches, say 554, the simulation shows a
maximum luminosity, which is also seen daily in the control room of the 
PEP-II. From the simulation, we see that the reason for the peaked 
luminosity is the rapid growth of $\sigma_y^+$ once the peak current 
is passed. 

In addition, the simulation predicts that we can reach the design 
luminosity 3$\times10^{33}{\rm cm}^{-2}{\rm s}^{-1}$ by running 829 
bunches at the beam current of $I_+ = 1600$mA and $I_- = 800$mA. This 
prediction has not been realized yet at this time. Currently, the total 
positron current is probably limited below 1200mA by the electron-cloud 
instability \cite{cloud}. Once this limitation is removed, we expect to 
reach the design luminosity with 829 bunches.

There is no particle loss outside the area ($8\sigma_x$$\times$$24\sigma_y$)
covered by the mesh in the first 15 data points. Beyond the 15th points, 
particle loss is almost about 1\%. 

\subsection{Damping Time}

\begin{figure}[ht]
 \vspace{9.0cm}
\includegraphics{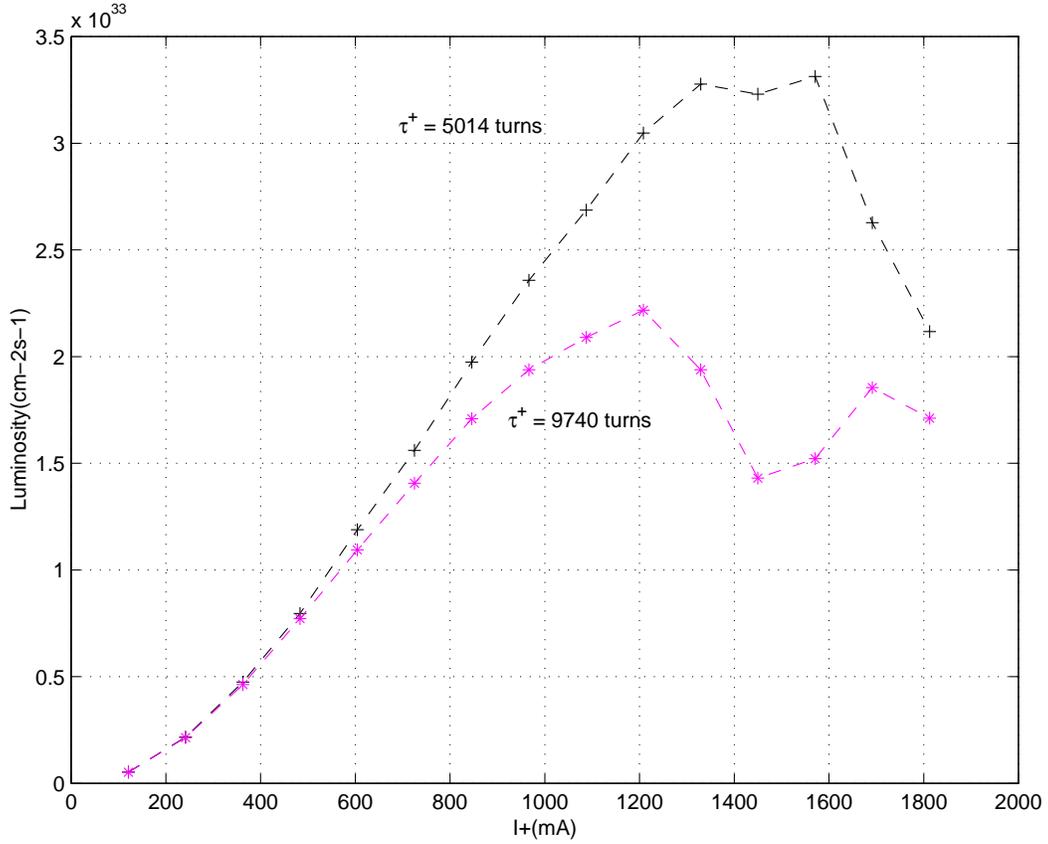}
 \vspace{3.5cm}
 \caption{\it Luminosity affected by the damping time with 554 bunches.} 
    \label{fig:tau} 
\end{figure}

Historically, the damping time is typically not considered to be an 
important parameter for the beam-beam effects. So we make an attempt to 
reduce the damping time artificially for the LER to speed up the 
computation. The result is shown in Fig.~\ref{fig:tau}

The only difference of the parameters used in two simulations is the
damping time in the LER, which is indicated as the labels in the figure.
Indeed, at the low current, the difference of the luminosity is rather
small, which is consistent with the simulation performed when the change of
the wiggler was made. But the difference grows larger, as the current 
increases. At the peak luminosity for the PEP-II operation, $I_+ = 1200$mA, 
the difference is about 40\%, which is significant.

This result shows for the first time that the damping time is a rather 
important parameter for the computation of the peak luminosity at high 
beam currents. Secondly, it points a way to improve the peak luminosity 
of the PEP-II without the increase of the beam currents, namely to install 
another wiggler in the LER to reduce the damping time to the original 
design value.

\renewcommand{\theequation}{\thesection.\arabic{equation}}

\setcounter{equation}{0}

\section{Discussion}

We have developed a hybrid method of solving the potential with an open 
boundary by using Green's function to fix the potential on a finite
boundary and then to solve the Poisson equation for the potential inside 
the boundary. The method is applied to the simulation of strong-strong 
interaction of beam-beam effects in PEP-II. The preliminary results 
of this simulation show a very good quantitative agreement with the 
experimental observations. Given the simplicity of the two-dimensional 
model used, the achievement is surprising and remarkable. We have 
demonstrated that the present code has a highly reliable predictive 
capability of realistic beam-beam interaction. To further benchmark the 
code, we need to extend the simulation to include the finite length of 
the bunch and compare the simulation results directly to the controlled 
experiments.

This method is quite general. It can be applied to the problem of space 
charge in three dimensions. It can also be used in the beam-beam interaction
of a linear collider. Finally, it can be applied to any boundary condition to
reduce the region of the mesh if Green's function is known.

\subsection*{Acknowledgments}

We would like to thank John Irwin, John Seeman and Ron Ruth for their
continuous support and encouragement. We would like also to thank 
Franz-Josef Decker, Miguel Furman, Sam Heifets, Albert Hoffmann, 
Witold Kozanecki, Michiko Minty, Robert Siemann, Mike Sullivan, 
Robert Warnock, Uli Wienands and Yiton Yan for the helpful discussions. 
Especially, we would like to thank Srinvas Krishnagopal for many 
explanations of the PIC method during his visit at SLAC. One of the
authors (TT) is supported in part by DOE contract W-7405-Eng.48 and
DOE grant DE-FG03-96ER40954.

\newpage

 

\end{document}